\documentclass{aa}  
 \usepackage{comment} 
\usepackage{tikz}
\usetikzlibrary{positioning, arrows.meta, calc, shapes.geometric}
\usepackage{booktabs} 
\usepackage{graphicx}
\usepackage{txfonts}
\usepackage{lipsum}
\usepackage{float}
\usepackage{longtable}
\usepackage{amssymb}
\usepackage{lineno}
\usepackage{booktabs}
\usepackage{multirow}
\usepackage{lscape}             
\usepackage{placeins}           
                                
\modulolinenumbers[1]

\usepackage{natbib}
\bibpunct{(}{)}{;}{a}{}{,} 
\usepackage{enumitem}
\begin{document}
    \title{Deep Learning-Based Classification and Analysis of Pulsar Candidates in \textit{Fermi}-LAT Unassociated Sources}
    \titlerunning{1D-CNN to classify and analysis pulsars in \textit{Fermi}-LAT Unassociated Sources}


   \author{C. Pozo González \inst{1}, R. López-Coto \inst{1}, J. Méndez-Gallego \inst{1}, E. de Oña Wilhelmi \inst{2}}

   \institute{Instituto de Astrofísica de Andalucía, CSIC, Glorieta de la Astronomía s/n, E-18008 Granada, Spain
   \and 
   Deutsches Elektronen-Synchrotron DESY, Platanenallee 6, 15738, Zeuthen, Germany
   }
   \authorrunning{C. Pozo González \inst{1}, et al}
   \date{}

 
  \abstract
  {The \textit{Fermi} Large Area Telescope (LAT) has revolutionized our understanding of the high-energy sky, yet approximately one-third of the sources in the Fourth \textit{Fermi}-LAT Source Catalog (4FGL) remain unassociated. Conventional machine learning, such as Decision Trees, often treat spectral features as independent tabular entries, neglecting the sequential topological information inherent in the Spectral Energy Distribution (SED). Additionally, they frequently rely on spatial coordinates, which introduces location bias.}
  {We aim to classify unassociated \textit{Fermi}-LAT sources by exploiting the intrinsic shape of their spectra and variability features, avoiding the use of galactic coordinates as training features. Our primary objective is to generate a high-confidence list of PSRs candidates, further distinguishing between Young Pulsars (YPs) and Millisecond Pulsars (MSPs)}
  {We developed a hierarchical deep learning framework based on a 1D Convolutional Neural Network (1D-CNN), named \textit{TabularResCNN}. This architecture treats the spectral data from the 4FGL catalog, allowing the model spectral shape. The classification is performed in two stages: first discriminating between AGNs and PSRs, and subsequently categorizing PSRs into YPs and MSPs. We implemented a cost-sensitive learning strategy to handle class imbalance and utilized Grad-CAM techniques to ensure the physical interpretability of the model's decisions.}
  {Applying this framework to 2563 unassociated sources, we identified 1136 AGNs and 202 high-confidence PSR candidates (166 YPs, 36 MSPs), increasing the pulsar population by more than 60\%. They exhibit strong astrophysical consistency: YPs are confined to the Galactic plane, MSPs show a broader vertical distribution, and AGNs are isotropic. Furthermore, we identified 5 out of 5 PSRs recently confirmed by FAST.}
  {The proposed 1D-CNN framework isolates PSRs candidates based on intrinsic spectral and temporal properties, minimizing spatial bias. The resulting catalog provides a robust set of targets for deep radio surveys with current and future facilities such as FAST, SKAO or CTAO.}


   \maketitle
   \nolinenumbers

\section{Introduction}
Since its launch in June 2008, the \textit{Fermi} Gamma-ray Space Telescope, equipped with the Large Area Telescope (LAT), has been continuously surveying the $\gamma$-ray sky, significantly expanding our knowledge of high-energy sources \citep{Atwood2009}. The \textit{Fermi}-LAT Collaboration periodically publishes all-sky source catalogs derived from accumulated data. The Fourth \textit{Fermi}-LAT Source Catalog (4FGL) initially detailed 5064 sources based on eight years of observations \citep{4FGLDR3}. Subsequent incremental releases, such as DR3 and DR4, have extended this dataset to over fourteen years, refining spectral models and increasing the population of PSRs and blazars.

Despite these catalog improvements, a significant subset of sources remains unassociated, lacking firm multi-wavelength counterparts. Proper classification of these objects is critical for understanding the demographics of the active galactic nuclei (AGNs) and pulsar (PSRs) populations and for prioritizing follow-up observations. Machine Learning (ML) has become the standard approach for this task. Early works applied Random Forest and logistic regression to previous catalogs, achieving high accuracies ($>95\%$) in binary classification \citep{SazParkinson2016, Zhu2020}. Later studies refined these benchmarks using gradient boosting techniques like CatBoost and ensemble methods to tackle more granular sub-classification tasks \citep{Coronado_Bl_zquez_2022, Zhu2023}.

However, as model complexity increases, methodological concerns regarding feature selection have emerged. For instance, \citet{2024MNRAS.527.1794Z} explicitly incorporated spatial coordinates (e.g., Galactic latitude) as dominant training features in their classification of 4FGL-DR3 sources. While this strategy yields high validation accuracy by exploiting the known concentration of YPs in the Galactic disk, it introduces a significant location bias. By conflating intrinsic astrophysical emission properties with line-of-sight coordinates, such models risk overfitting to the telescope's selection function and the structured Galactic background. Consequently, this approach compromises the ability to detect outlier populations, such as high-latitude MSPs or halo candidates, which do not conform to the spatial priors of the training set.

More recently, \citet{2026APh...17503185P} established a state-of-the-art benchmark on the 4FGL-DR4 catalog using XGBoost and Random Forest. Their work emphasizes the role of data balancing in addressing class disparity. However, while their results are statistically robust, tree-based models treat spectral and temporal properties strictly as independent tabular features.

Consequently, they are permutation invariant effectively ignoring the sequential order of spectral bins, and struggle to capture complex, non-linear interdependencies without extensive manual feature engineering. Furthermore, the aggressive synthetic oversampling often employed in these studies risks introducing statistical artifacts that do not reflect the physical reality of faint sources.

To transcend these limitations, we propose that the classification of $\gamma$-ray sources requires models capable of exploiting the intrinsic topological structure of the data. In spectroscopic analysis, input parameters are not isolated entities; the flux intensity in a specific energy band is physically correlated with its neighboring bands. Convolutional Neural Networks (CNNs), particularly 1D-CNNs, provide a distinct advantage here. Unlike decision trees, 1D-CNNs utilize convolutional kernels to scan local neighborhoods, inherently learning shape-dependent features (such as spectral curvature, cutoffs, and breaks) directly from the relative positions and gradients of adjacent input parameters.

In this work, we introduce a two-stage 1D-CNN deep learning framework tailored for tabular $\gamma$-ray features from the 4FGL-DR4 catalog. The first stage discriminates AGNs from PSRs, and the second further separates YPs from MSPs. To mitigate class imbalance, we applied class weighting to the loss function when necessary. Furthermore, we used interpretability techniques (e.g., Grad-CAM) to highlight relevant physical features, and evaluated the model's performance on both associated and unassociated sources.

\section{Data and Preprocessing}
\label{sec:data}

This study utilizes the most recent $\gamma$-ray source catalog released by the \textit{Fermi}-LAT collaboration, complemented by radio PSRs data for validation purposes.

\subsection{The \textit{Fermi}-LAT 4FGL-DR4 Catalog}
The primary dataset is derived from the Fourth \textit{Fermi}-LAT Source Catalog, Data Release 4 (4FGL-DR4)\footnote{Available at: \url{https://fermi.gsfc.nasa.gov/ssc/data/access/lat/14yr_catalog/}}, based on 14 years of observations in the 50\,MeV to 1\,TeV energy range. The catalog contains 7195 $\gamma$-ray sources, characterized by 79 distinct features including spectral parameters, spatial morphology, and variability indices.

The catalog is dominated by AGNs, specifically Blazars (BLL, FSRQ, and BCU), which constitute approximately 54.7\% of the population. PSRs represent a minority class, accounting for only 4.45\% of the associated sources. A significant portion of the catalog (33.7\%) remains unassociated, lacking firm counterparts at other wavelengths. Table \ref{tab:catalog_dist} summarizes the class distribution relevant to this work.

\begin{table*}
    \centering
    \caption{Distribution of source classes in the 4FGL-DR4 catalog.}
    \label{tab:catalog_dist}
    \small
    \begin{tabular}{llrr}
        \toprule
        \textbf{Class} & \textbf{Description} & \textbf{Sources} & \textbf{Percentage} \\
        \midrule
        \multicolumn{4}{l}{\textit{Training \& Validation Classes}} \\
        \midrule
        BLL  & BL Lac object & 1,468 & 20.40\% \\
        FSRQ & Flat Spectrum Radio Quasar & 776 & 10.79\% \\
        BCU  & Blazar of Uncertain type & 1,622 & 22.54\% \\
        \cmidrule(lr){1-4}
        \textbf{AGNs Total} & \textbf{(BLL + FSRQ + BCU)} & \textbf{3,866} & \textbf{53.73\%} \\
        
        \midrule
        YPs & Young Pulsar & 141 & 1.96\% \\
        MSPs & Millisecond Pulsar & 179 & 2.49\% \\
        \cmidrule(lr){1-4}
        \textbf{PSRs Total} & \textbf{(PSR + MSP)} & \textbf{320} & \textbf{4.45\%} \\
        
        \midrule
        \multicolumn{4}{l}{\textit{Target for Prediction}} \\
        \midrule
        UNK & Unassociated Sources & 2,423 & 33.68\% \\
        \bottomrule
    \end{tabular}
\end{table*}

\subsection{The ATNF PSRs Catalog}
To provide physical significance to our validation datasets, we cross-matched our sample with the Australia Telescope National Facility (ATNF) PSRs Catalogue (v2.6.3).\footnote{Available at: \url{https://www.atnf.csiro.au/research/pulsar/psrcat/}}. This catalog contains $4346$ known radio PSRs and provides crucial timing parameters, specifically the spin period ($P$) and period derivative ($\dot{P}$), which are not available in the \textit{Fermi}-LAT $\gamma$-ray data.

\subsection{Training Dataset Construction}
The 4FGL-DR4 catalog sources will be structured into two classification levels based on their physical and population properties. The primary level will comprise $4261$ sources, divided into AGNs ($N=3941$), encompassing BLL, FSRQ, and BCU subtypes, and PSRs ( $N=320$), which will include both young and millisecond varieties. This distribution will exhibit a class imbalance of approximately 12:1. Data allocation will follow an $80\%$ stratified split for training and $20\%$ for validation, where the asymmetry will be managed via a cost-sensitive learning strategy. This approach will apply inverse class weights to the cross-entropy loss function to penalize misclassification of the minority class without introducing synthetic artifacts.

Within the PSRs population, the sample will be categorized into 141 YPs ($P > 30$\,ms) and 179 MSPs ($P < 30$\,ms), representing 44\% and 56\% of the subset, respectively. Given the relative equilibrium of this $\sim1:1.2$ ratio, the model will learn directly from the original spectral distributions of confirmed sources, ensuring the capture of authentic physical features. 

Finally, a separate set of $2423$ unassociated sources will be reserved exclusively for the inference phase following the model validation process.

\section{Methodology}
\label{sec:methodology}

This section details the end-to-end Deep Learning framework developed for the classification of unassociated \textit{Fermi}-LAT sources. Our approach prioritizes the identification of candidate PSRs through a hierarchical architecture that takes advantage of both the spectral and temporal structure of the 4FGL-DR4 data.

\subsection{Feature Engineering}
\label{sec:features_engineering}
The primary dataset is derived from the \textit{Fermi}-LAT 4FGL-DR4 catalog \citep{4FGLDR3}. The raw FITS data is processed to construct a high-dimensional feature space suitable for convolutional analysis.

We extract a comprehensive set of parameters characterizing the $\gamma$-ray emission:
\begin{itemize}
    \item Spectral Features: Integrated flux, energy flux per band, spectral index, pivot energy, and spectral curvature significance.
    \item Variability Features: Variability Index ($TS_{var}$) and fractional variability, which are critical for distinguishing stochastic AGNs flaring from stable PSR emission.
    \item Detectability: Detection significance and $\sqrt{TS}$ across individual energy bands.
\end{itemize}

A critical step in our pipeline is the Array Expansion. Multi-dimensional FITS columns, such as \texttt{Flux\_Band} (spanning 7 energy bands), are flattened and sequentially arranged. This preserves the energy-dependent ordering, allowing the subsequent CNN layers to interpret the SED as a coherent 1D signal rather than a set of disjoint features.

Astronomical data often contain missing values or upper limits due to varying sensitivity across the sky. Standard imputation (e.g., mean substitution) destroys this information. To address this, we implement a Dual-Channel Input Strategy:
\begin{itemize}
    \item Channel 1 (Value Map): Contains the normalized numerical features. Missing entries are zero-filled to maintain input tensor shape.
    \item Channel 2 (Sensitivity Mask): A binary tensor where $1$ denotes a valid measurement and $0$ denotes a missing value or upper limit.
\end{itemize}
This architecture allows the network to explicitly learn the patterns of observability, distinguishing between a source that is physically faint (low flux) and one that is unobserved due to instrumental limitations.

\subsection{Hierarchical Classification Framework}
Given the class imbalance and the distinct physical nature of the background sources, we eschew a monolithic multi-class classifier in favor of a Hierarchical Stage-wise Approach:

\begin{enumerate}[label=\textbf{Stage \arabic*:}, leftmargin=*]
    \item Source Class Discrimination (AGNs vs. PSRs). \\
    A binary classifier separates sources into AGNs and PSRs. This segregation exploits the primary dichotomy in the $\gamma$-ray sky: AGNs typically exhibit high variability and softer spectra, while PSRs are generally stable with curved spectra (cutoff in the GeV range).
    
    \item PSR Sub-typing (YPs vs. MSPs). \\
    Sources classified as PSRs with high confidence ($P_{PSR} > 0.5$) are passed to a specialized secondary model. This stage distinguishes between YPs and MSPs, characterized by higher magnetic fields and slower rotation, and MSPs, which typically exhibit harder spectra and different orbital demographics.
\end{enumerate}

\subsection{Model Architecture: \textit{TabularResCNN}}
The core of our classification engine is the \textit{TabularResCNN}, a custom 1D Residual Convolutional Neural Network designed to extract local correlations from tabular data. 

Unlike fully connected networks, which treat inputs as permutation-invariant, \textit{TabularResCNN} respects the topological structure of the expanded feature vector. The architecture consists of:
\begin{itemize}
    \item 1D Convolutional Layers: Filters slide across the energy dimension of the feature vector, detecting local gradients (e.g., spectral breaks) and inter-band correlations.
    \item Residual Blocks: We employ skip connections ($y = F(x) + x$) to facilitate gradient flow, allowing the model to learn deep hierarchical representations without degradation.
    \item Global Average Pooling (GAP): Following feature extraction, GAP aggregates the feature maps across the sequence dimension. This makes the model robust to minor shifts in feature ordering while reducing the parameter count, thereby mitigating overfitting.
    \item Classification Head: A multi-layer perceptron (MLP) maps the aggregated features to the final probability score.
\end{itemize}
In the Appendix \ref{app:architecture} you can check all the hyperparameters and the architectural designs in more detail.

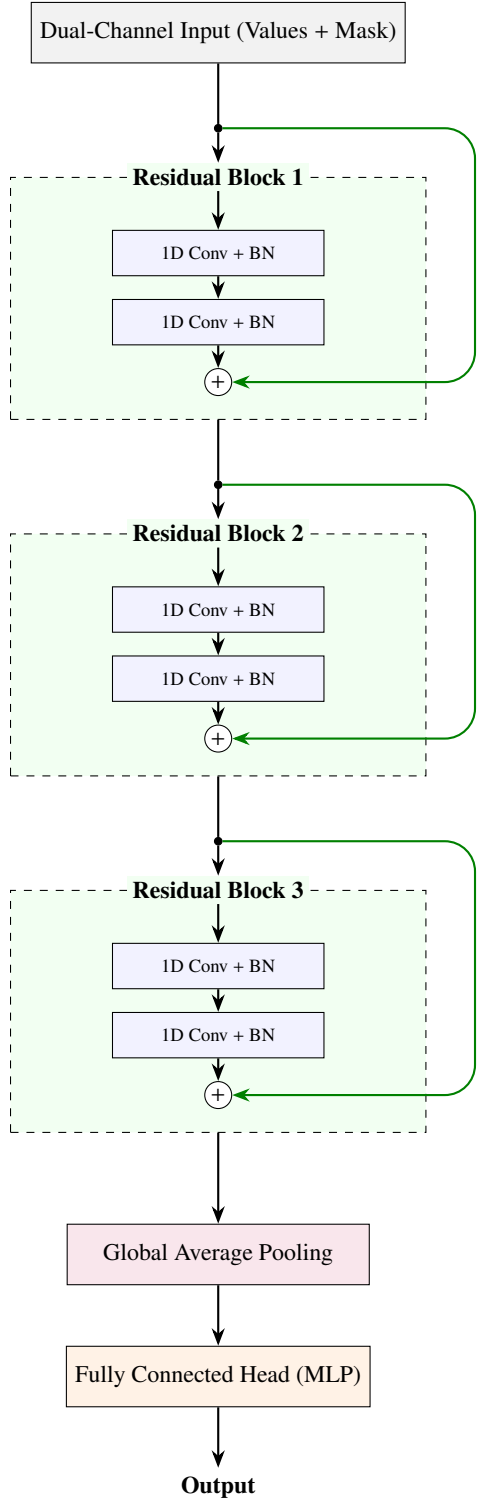
\begin{figure}[ht] 
\centering
\begin{tikzpicture}[
    node distance=0.8cm and 2.2cm,
    layer/.style={rectangle, draw, fill=blue!10, text centered, minimum height=0.8cm, minimum width=4cm, font=\small},
    resblock/.style={rectangle, draw, fill=green!5, minimum height=3.2cm, minimum width=5.5cm, dashed, inner sep=0.3cm},
    blocktitle/.style={fill=green!5, inner sep=2pt, font=\small\bfseries},
    sublayer/.style={rectangle, draw, fill=blue!5, text centered, minimum height=0.6cm, minimum width=2.8cm, font=\scriptsize},
    plus/.style={circle, draw, fill=white, inner sep=1pt, font=\small},
    branch/.style={circle, draw, fill=black, inner sep=1pt},
    arrow/.style={-Stealth, thick},
    line/.style={thick}, 
    skip_u/.style={-Stealth, thick, color=green!50!black, rounded corners=4mm}
]
    
    \node (input) [layer, fill=gray!10] {Dual-Channel Input (Values + Mask)};

    \node (b1_branch) [branch, below=0.8cm of input] {}; 
    
    \node (block1) [resblock, below=1.5cm of input] {};
    \node (title1) at (block1.north) [blocktitle, anchor=center] {Residual Block 1};
    \node (conv1a) [sublayer, shift={(0, 0.6)}, at={(block1.center)}] {1D Conv + BN};
    \node (conv1b) [sublayer, below=0.3cm of conv1a] {1D Conv + BN};
    \node (plus1) [plus, below=0.3cm of conv1b] {$+$};

    \draw [line] (input) -- (b1_branch);
    \draw [arrow] (b1_branch) -- (title1.north);
    \draw [arrow] (title1.south) -- (conv1a.north);
    
    \draw [arrow] (conv1a) -- (conv1b);
    \draw [arrow] (conv1b) -- (plus1);
    
    \draw [skip_u] (b1_branch) -- ++(3.4, 0) |- (plus1.east);

    \node (b2_branch) [branch, below=0.8cm of block1] {}; 

    \node (block2) [resblock, below=1.5cm of block1] {};
    \node (title2) at (block2.north) [blocktitle, anchor=center] {Residual Block 2};
    \node (conv2a) [sublayer, shift={(0, 0.6)}, at={(block2.center)}] {1D Conv + BN};
    \node (conv2b) [sublayer, below=0.3cm of conv2a] {1D Conv + BN};
    \node (plus2) [plus, below=0.3cm of conv2b] {$+$};

    \draw [line] (block1.south) -- (b2_branch);
    \draw [arrow] (b2_branch) -- (title2.north);
    \draw [arrow] (title2.south) -- (conv2a.north);
    
    \draw [arrow] (conv2a) -- (conv2b);
    \draw [arrow] (conv2b) -- (plus2);

    \draw [skip_u] (b2_branch) -- ++(3.4, 0) |- (plus2.east);

    \node (b3_branch) [branch, below=0.8cm of block2] {}; 

    \node (block3) [resblock, below=1.5cm of block2] {};
    \node (title3) at (block3.north) [blocktitle, anchor=center] {Residual Block 3};
    \node (conv3a) [sublayer, shift={(0, 0.6)}, at={(block3.center)}] {1D Conv + BN};
    \node (conv3b) [sublayer, below=0.3cm of conv3a] {1D Conv + BN};
    \node (plus3) [plus, below=0.3cm of conv3b] {$+$};

    \draw [line] (block2.south) -- (b3_branch);
    \draw [arrow] (b3_branch) -- (title3.north);
    \draw [arrow] (title3.south) -- (conv3a.north);
    
    \draw [arrow] (conv3a) -- (conv3b);
    \draw [arrow] (conv3b) -- (plus3);

    \draw [skip_u] (b3_branch) -- ++(3.4, 0) |- (plus3.east);

    \node (gap) [layer, below=1.2cm of block3, fill=purple!10] {Global Average Pooling};
    \node (fc) [layer, below=0.8cm of gap, fill=orange!10] {Fully Connected Head (MLP)};
    \node (output) [below=0.8cm of fc, font=\small\bfseries] {Output};

    \draw [arrow] (block3.south) -- (gap);
    \draw [arrow] (gap) -- (fc);
    \draw [arrow] (fc) -- (output);

\end{tikzpicture}
\caption{Architecture of the \textit{TabularResCNN} model. Main flow enters the residual block boundary before processing.}
\label{fig:architecture}
\end{figure}

\subsection{Training and Optimization}
The models are trained using the Adam optimizer with a dynamic learning rate scheduler (\texttt{ReduceLROnPlateau}), which decays the learning rate upon validation loss stagnation. To enhance generalization:
\begin{itemize}
    \item Loss Function: We utilize Cross-Entropy Loss with label smoothing ($\epsilon=0.1$). This technique prevents the model from becoming over-confident in training labels, which is crucial given the potential for misclassified labels in the training catalogs.
    \item Validation Strategy: We employ a stratified 80/20 train-test split to preserve the natural class distribution, ensuring that rare classes (e.g., PSRs) are adequately represented in the evaluation set.
\end{itemize}

\subsection{Class Imbalance Mitigation}
\label{sec:imbalance_strategy}

The training dataset exhibits a significant class imbalance in stage 1, with AGNs outnumbering PSRs by a ratio of approximately $12:1$. While traditional machine learning workflows frequently employ synthetic augmentation, such as SMOTE or ADASYN, to mitigate such disparities, we intentionally bypassed these techniques in favor of preserving the physical integrity of the sample.We contend that within high-energy astrophysical catalogs, synthetic oversampling introduces a fundamental methodological risk: the amplification of detection bias. As noted by \cite{Malyshev_2025}, these sources represent only the brightest and most spectrally distinct objects accessible to current instrumentation.

Synthetic generation algorithms operate by interpolating between existing minority samples in the feature space. Consequently, applying SMOTE to this specific dataset would artificially populate the manifold of bright and clear PSRs, reinforcing a decision boundary that is biased towards high-significance sources. This is counterproductive for our primary objective: the classification of unassociated sources. These unassociated targets are typically fainter, located in complex backgrounds, or possess lower detection significance than the training set.

Therefore, training on synthetically augmented data would likely induce a covariate shift, where the training distribution $P_{train}(X)$ diverges from the target inference distribution $P_{target}(X)$. To mitigate this without compromising physical fidelity, we adopted a Cost-Sensitive Learning strategy.

We implemented a Weighted Cross-Entropy Loss function, where the contribution of each sample to the gradient is scaled inversely proportional to its class frequency. The weight $w_j$ for class $j$ is defined as:
\begin{equation}
    w_j = \frac{N_{total}}{2 \cdot N_j}
\end{equation}
where $N_{total}$ is the total number of samples and $N_j$ is the count of samples in class $j$. This forces the optimizer to penalize the misclassification of a single real PSR significantly more than that of an AGNs, effectively balancing the learning dynamics while ensuring the model learns exclusively from authentic, albeit sparse, physical data.

\subsection{Cost-sensitive threshold optimization}
\label{sec:methods_thresholds}

In standard binary classification tasks, the decision boundary is typically set at a default probability threshold of $P_{AGNs} = P_{PSRs} = 0.5$. However, evaluating model performance using metrics that treat false positives and false negatives symmetrically (such as accuracy or F1-score) fails to capture the operational reality of astronomical follow-up campaigns. In our scientific context, the observational penalty of misclassifying a background fluctuation as a PSR  fundamentally differs from the scientific penalty of missing a true PSR. 

To address this asymmetry, we implemented a cost-sensitive threshold optimization strategy. Rather than blindly maximizing generic mathematical metrics, we shifted the paradigm from mathematical optimization to observational resource allocation. We evaluated the decision boundaries using observational efficiency curves (analogous to lift curves; see Appendix \ref{app:efficiency}), which quantify the expected target yield as a function of the candidate ranking against a random selection baseline.

To mitigate the risk of threshold overfitting, a common pitfall where the optimal probability cutoff becomes heavily biased by the specific statistical fluctuations of a limited validation set, we employed a bootstrapping technique. We generated $1000$ random resamples (with replacement) of our validation dataset's probability predictions. For each bootstrap iteration, we calculated the threshold that maximized our predefined observational efficiency. 

The final operational thresholds applied to the unassociated catalog were derived from the median of this bootstrapped distribution. Consequently, the network dynamically adapts its strictness: imposing stringent, high-purity boundaries for highly separable classes, while adjusting boundaries for classes that blend with the background to guarantee the creation of high-confidence candidate lists.

\subsection{Model Interpretability}

To identify which variables most influence the model’s predictions, the Grad-CAM technique \citep{Selvaraju_2019} was applied.  
This method computes the gradients of the output with respect to internal feature maps, enabling visualization of how each parameter contributes to the final classification.  
The resulting activation maps indicate that variables such as average significance, spectral index, and variability are particularly important for distinguishing between PSRs and AGNs.  
This approach provides a physically interpretable link between the machine-learning model and the underlying astrophysical properties.

\subsection{Verification}
To validate the astrophysical reliability of our candidates, we implement a post-inference analysis pipeline:

\begin{enumerate}
    \item Spatial Cross-Matching: High-confidence candidates are cross-referenced with the ATNF PSRs Catalog. Matches are defined not by a simple radius, but by the intersection of the ATNF coordinates with the \textit{Fermi}-LAT $95\%$ confidence error ellipses.
    \item P-Pdot Diagram Validation: For candidates associated with known radio PSRs, we plot their spin period ($P$) versus period derivative ($\dot{P}$). This serves as a physical sanity check to confirm that sources classified as MSPs dynamically reside in the recycled PSR region of the diagram.
    \item External Validation (FAST Discoveries): The ultimate test of the methodology is its predictive power on unseen data. We benchmark our unassociated source candidates against recent blind-search discoveries from the Five-hundred-meter Aperture Spherical Radio Telescope (FAST), verifying if our model correctly flagged these sources prior to their radio confirmation.
\end{enumerate}

\section{Results}
\label{sec:results}

We first evaluate the statistical performance of the model on the labelled test set, followed by a detailed analysis of the physical properties and spatial distribution of the newly identified candidates.

\subsection{Statistical Performance Assessment}
We first evaluate the pure discriminative power of the hierarchical framework on the labeled validation set. To establish the model's peak mathematical performance before applying observational constraints, we derived the optimal probability thresholds that maximize the overall accuracy for each stage. In this binary context, where the probabilities are complementary ($P_{AGNs} = 1 - P_{PSRs}$), a source is classified as an AGNs if $P_{AGNs} \geq 0.53$, and as a PSR if $P_{PSRs} > 0.46$. Evaluated at these boundaries, the metrics summarized in Tables \ref{tab:stage1_metrics} and \ref{tab:stage2_metrics} indicate that the model effectively disentangles the spectral degeneracies inherent in $\gamma$-ray sources.

In the first stage, tasked with separating AGNs from PSRs, the model demonstrates exceptional sensitivity and an overall accuracy of 96.7\%. As detailed in Table \ref{tab:stage1_metrics}, the performance is characterized by distinct behaviors for each class:
\begin{itemize}
    \item AGNs: The network achieves near-perfect retrieval and purity for extragalactic sources. This confirms that the stochastic flaring and softer power-law spectra characteristic of blazars are highly distinctive features successfully captured by the 1D-CNN.
    \item PSRs: While the classifier maintains high precision for the combined PSRs class, it exhibits a comparatively lower recall. This relative drop is an expected physical limitation.
\end{itemize}

The second stage, which distinguishes between YPs and MSPs, presents a significantly more complex challenge due to the intrinsic spectral and physical overlap between these two neutron star populations. Nevertheless, the model maintains a robust discriminative capacity (Table \ref{tab:stage2_metrics}):
\begin{itemize}
    \item MSPs: The classifier proves highly effective at isolating MSPs, yielding strong precision and a well-balanced F1-Score that outperforms the YP classification.
    \item YPs: The performance on Young Pulsars reflects the inherent difficulty of this granular task. Misclassifications in this stage are primarily driven by older YPs entering the transition phase, or extremely faint sources where the high-energy exponential cutoff, a key discriminant for young pulsars, is not robustly resolved in the observations.
\end{itemize}

\begin{table*}
    \centering
    \caption{Performance Metrics for Stage 1: AGNs vs. Pulsar Candidates (Overall Accuracy: 96.7\%)}
    \label{tab:stage1_metrics}
    \small
    \begin{tabular}{lcccc}
        \toprule
        \textbf{Class} & \textbf{Precision} & \textbf{Recall} & \textbf{F1-Score} & \textbf{Support} \\
        \midrule
        AGNs & 0.9853 & 0.9937 & 0.9895 & 473 \\
        PSRs & 0.9500 & 0.8906 & 0.9194 & 64 \\
        \midrule
        \textbf{Macro Average} & \textbf{0.9677} & \textbf{0.9421} & \textbf{0.9544} & \textbf{537} \\
        \bottomrule
    \end{tabular}
\end{table*}

\begin{table*}
    \centering
    \caption{Performance Metrics for Stage 2: Young PSR vs. MSP (Overall Accuracy: 81.4\%)}
    \label{tab:stage2_metrics}
    \small
    \begin{tabular}{lcccc}
        \toprule
        \textbf{Class} & \textbf{Precision} & \textbf{Recall} & \textbf{F1-Score} & \textbf{Support} \\
        \midrule
        MSPs & 0.8824 & 0.8108 & 0.8451 & 37 \\
        PSRs & 0.7200 & 0.8182 & 0.7660 & 22 \\
        \midrule
        \textbf{Macro Average} & \textbf{0.8012} & \textbf{0.8145} & \textbf{0.8055} & \textbf{59} \\
        \bottomrule
    \end{tabular}
\end{table*}

\subsection{New $\gamma$-ray candidates}
While the mathematically optimal thresholds derived in the previous section ($P_{AGNs} = 0.53$ and $P_{PSRs} = 0.46$ ) establish the peak discriminative power of the network, applying them to generate a target catalog is observationally sub-optimal. For context, applying standard symmetric-like cuts (e.g., a default $0.5$) to the unassociated sample yields 1514 AGNs and 1049 PSR candidates. This rigid mathematical approach needlessly contaminates the highly separable AGNs sample with background fluctuations, while burdening the pulsar candidate list with low-confidence sources that would heavily dilute the efficiency of radio follow-up campaigns.To address this, we applied the physically motivated cuts derived from our observational efficiency analysis (see Section \ref{sec:methods_thresholds} and Appendix \ref{app:efficiency}). By enforcing our optimized, cost-sensitive thresholds, specifically $P(\mathrm{AGNs}) \geq 0.86$ to guarantee high purity, and $P(\mathrm{PSR}) \geq 0.96$ to maximize observational efficiency for radio follow-ups, the application of the trained hierarchy to the 2563 unassociated sources in the 4FGL-DR4 catalog resulted in the classification distribution shown in Figure \ref{fig:latitude_dist}. The model identifies:

\begin{itemize}
    \item 1136 AGNs candidates ($44.3\%$). The strict probability cut ensures a highly pure catalog, filtering out 378 marginal sources that a 0.5 threshold would have loosely classified as AGNs.
    
    \item 202 PSRs candidates ($7.9\%$), significantly expanding the potential census of $\gamma$-ray PSRs. This subset is further resolved into 166 YPs and 36 MSPs. By elevating the decision boundary to prioritize sample purity over absolute completeness, we systematically filter out 847 low-confidence candidates (compared to the baseline 0.5 cut). This conservative strategy forms a high-confidence catalog, ensuring that multi-wavelength resources are exclusively allocated to the most robust targets.
\end{itemize}

\subsection{Spatial Analysis and Galactic Structure}
A primary validation of our results is the spatial distribution of the candidates. Since the model was trained primarily on spectral and temporal features, the emergence of coherent galactic structures in the predictions serves as strong independent evidence of physical validity.

\begin{figure*}[t]
    \centering
    \includegraphics[width=\textwidth]{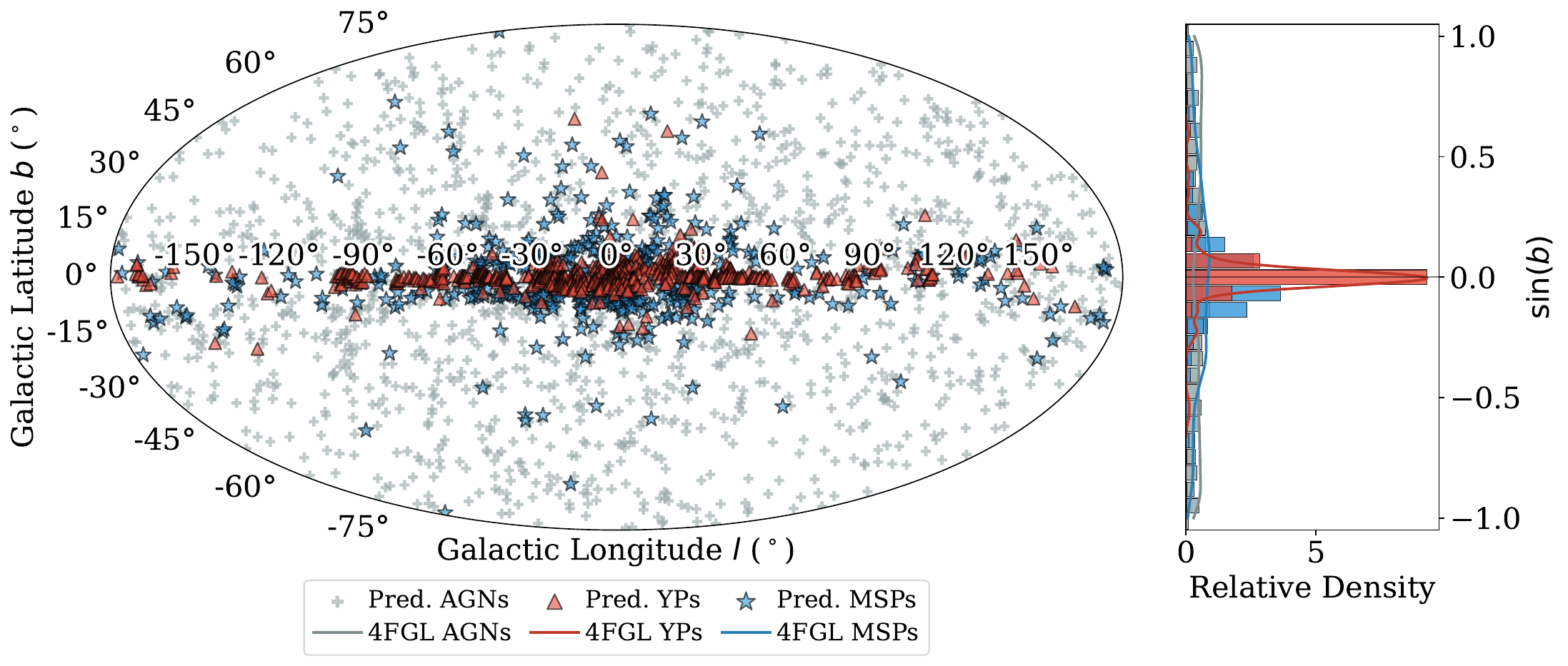}
    \caption{Spatial distribution of classified candidates. Left: Sky map in Galactic coordinates. Right: Marginal histogram of $\sin(b)$ solid and the 4FGL dashed. Note the tight confinement of Young PSRs to the plane versus the broader distribution of MSPs.}
    \label{fig:latitude_dist}
\end{figure*}

Figure \ref{fig:latitude_dist} presents the distribution of candidates as a function of Galactic sine latitude ($\sin b$).
\begin{itemize}
    \item The predicted YPs (red triangles) are strictly confined to the thin Galactic disk ($|b| < 5^\circ$), exhibiting a sharp peak at $\sin b \approx 0$.
    \item The MSP candidates (blue stars) show a broader vertical distribution, extending to intermediate latitudes. This puffed-up distribution is consistent with the kinematic heating of older stellar populations and the kick velocities imparted during supernova formation in binary systems.
    \item In contrast, the AGNs candidates (grey dots) display an isotropic distribution across the high-latitude sky, as expected for an extragalactic population.
\end{itemize}

\subsection{$P-\dot{P}$ Diagram}
Although the spin period ($P$) and period derivative ($\dot{P}$) are not inputs to our model (as they are unknown for unassociated sources), we performed a cross-match validation. We successfully associated 112 predictions with known PSRs in the ATNF catalog that have measured timing parameters.

\begin{figure}
    \centering
    \includegraphics[width=0.5\textwidth]{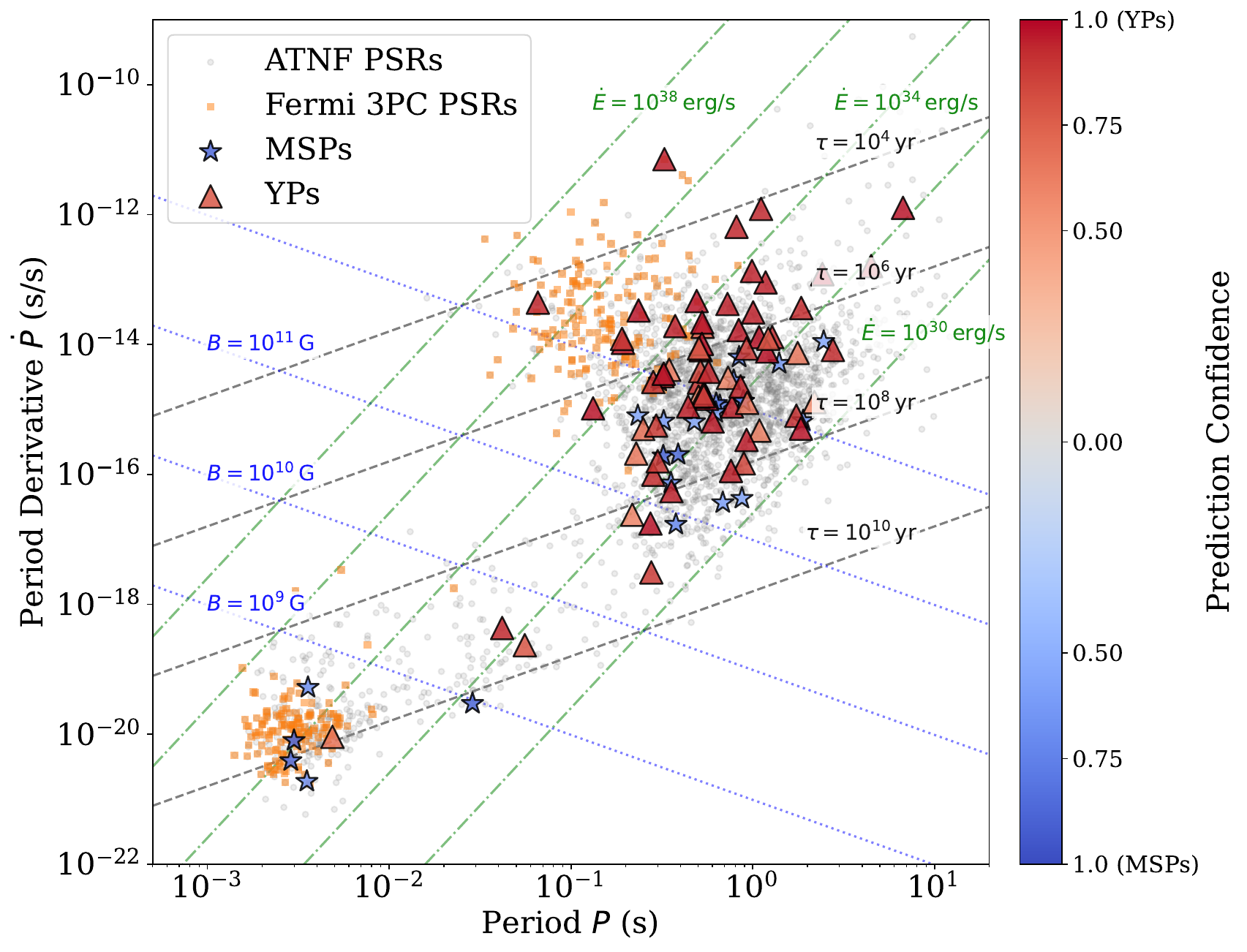}
    \caption{Pulsar candidates for which we have a cross-match with ATNF PSRs in the  diagram $P - \dot{P}$. Sources classified by our model as PSRs with  $P_{PSRs}>0.46$  in stage 1 and successfully cross-matched with the ATNF Pulsar Catalogue are shown. YPs are represented by triangles, while MSPs are indicated by star symbols. Marker colors encode the classification confidence assigned by the network. The background grey distribution corresponds to the full ATNF pulsar population, and gamma-ray PSRs from the Third \textit{Fermi}-LAT Catalog of Gamma-ray PSRs (3PC) are also displayed with orange triangles, for reference. Although timing parameters were not used during training, the model naturally recovers the two distinct physical populations in the plane, demonstrating that it implicitly captures underlying pulsar evolutionary properties.}
    \label{fig:ppdot}
\end{figure}

Figure \ref{fig:ppdot} shows these sources on the canonical $P-\dot{P}$ diagram. The results show a clear physical separation:
\begin{itemize}
    \item Sources classified as Young PSRs cluster in the upper-right island ($P \sim 0.1-1$\,s, $\dot{P} \sim 10^{-15}$), characteristic of isolated, rotation-powered PSRs.
    \item Sources classified as MSPs populate the lower-left region ($P < 10$\,ms, low magnetic fields), aligning perfectly with the recycled pulsar population.
\end{itemize}
This result confirms that the $\gamma$-ray spectral features learned by the \textit{TabularResCNN} are direct proxies for the intrinsic magnetospheric physics of the neutron stars.

\subsection{External Verification}
As a final test of predictive reliability, we cross-referenced our candidate list with the recent pulsar discoveries reported by the FAST (Five-hundred-meter Aperture Spherical Radio Telescope) survey. Our model correctly classified 5 out of 5 newly confirmed PSRs (Table \ref{tab:fast_candidates}) as high-confidence pulsar candidates prior to their radio confirmation. This 100\% \ agreement with blind radio searches underscores the utility of our catalog to prioritize future observational campaigns.

\begin{table*}[htbp]
\centering
\caption{FAST Pulsar Candidates and predictions of our model. Where $\text{Flux}_{1\,\rm GeV} = \int_{1\,\text{GeV}}^{\infty} \frac{dN}{dA \, dt \, dE} \, dE $}
\label{tab:fast_candidates}
\begin{tabular}{lccccccc}
\hline\hline
Source Name & $\ell$ & $b$ & Class & $P_{\rm PSRs}$ & $P_{\rm MSP}$ & $\sigma$ & Flux$_{1\,\rm GeV}$ \\
            & (deg)  & (deg) &       &              &       &          & ($10^{-11}$ ph cm$^{-2}$ s$^{-1}$) \\
\hline
4FGL J0237.8$+$5238  & 138.83 & -6.92 & MSP & 0.98 & 0.85 & 26.9 & 129.5 \\
4FGL J0533.6$+$5945  & 152.26 & 14.17 & MSP & 0.92 & 0.84 & 17.6 & 70.0 \\
4FGL J1730.4$-$0359  & 19.74 & 16.00 & MSP & 0.98 & 0.86 & 21.6 & 142.5 \\
4FGL J1827.5$+$1141  & 40.75 & 10.55 & MSP & 0.93 & 0.94 & 18.9 & 94.9 \\
4FGL J1904.7$-$0708  & 28.07 & -6.20 & PSR & 0.85 & 0.31 & 24.9 & 184.7 \\
\hline
\end{tabular}
\end{table*}

\section{Interpretability}
\label{sec:interpretability}

To ensure that the high performance of our 1D-CNN is not driven by statistical artifacts, we employed Gradient-weighted Class Activation Mapping (Grad-CAM). This technique allows us to visualize which spectral and temporal features strictly influence the model's decision-making process. The resulting activation maps demonstrate that the network has autonomously learned well-established astrophysical properties of $\gamma$-ray sources.

\subsection{Stage 1: AGNs vs PSRs}
\begin{figure*}
    \centering
    \includegraphics[width=1.\textwidth]{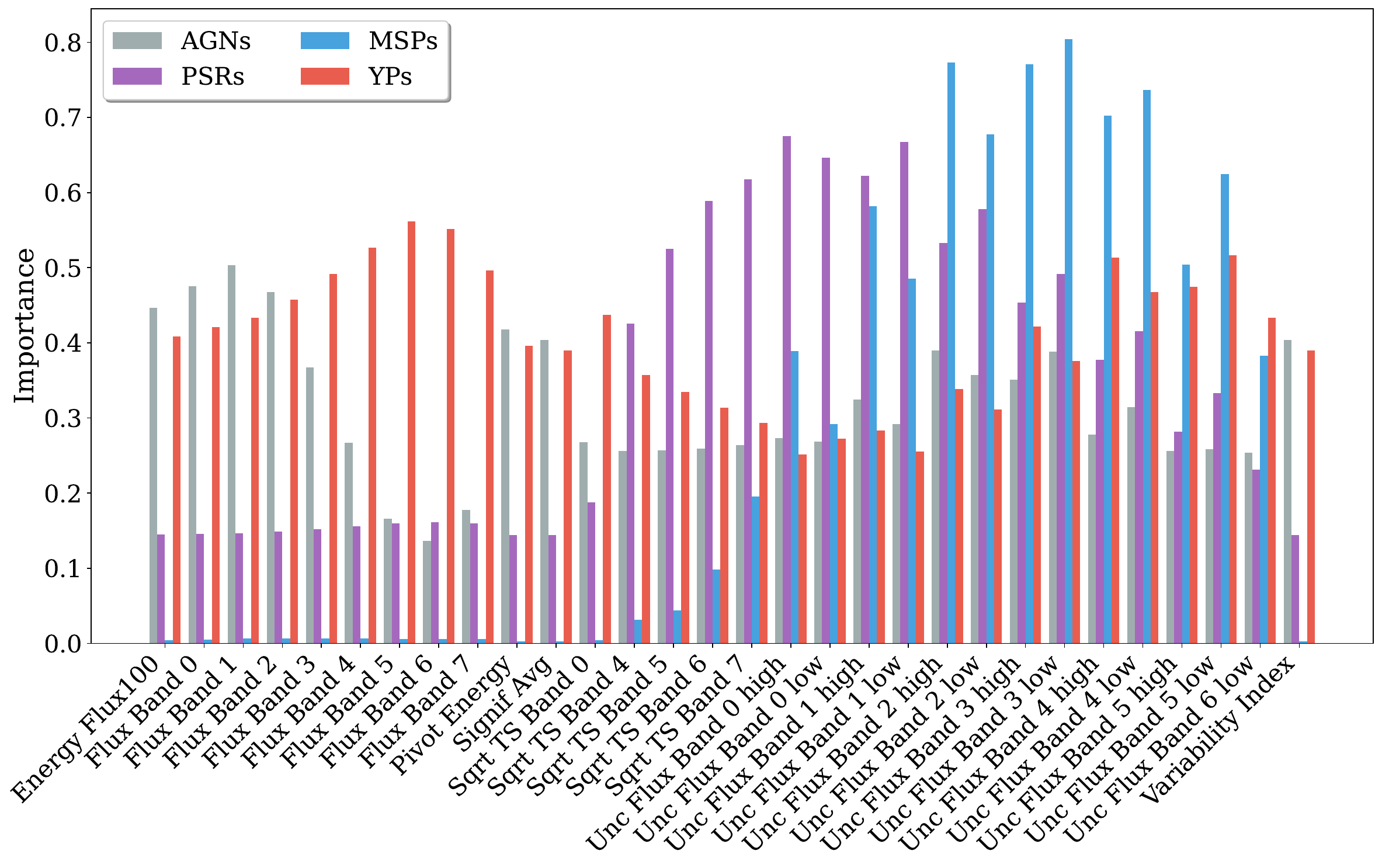}
    \caption{Average Grad-CAM feature importance for Stage 1 and 2}
    \label{fig:gradcam_stage1}
\end{figure*}

The discrimination between AGNs and PSRs sources reveals a clear physical bifurcation in feature importance (see Figure \ref{fig:gradcam_stage1}):

\begin{itemize}
    \item AGNs Identification: The Grad-CAM analysis for the AGNs class (Fig. \ref{fig:gradcam_stage1}) is dominated by low-energy flux components, specifically \texttt{Flux\_Band\_0} ($50-100$\,MeV) and \texttt{Flux\_Band\_1} ($100-300$\,MeV). Additionally, the \texttt{Variability\_Index} appears as a top-tier feature. This perfectly aligns with the physical nature of blazars (FSRQs and BL Lacs), which typically exhibit softer power-law spectra and stochastic flaring activity compared to PSRs.
    
    \item PSRs Identification: Conversely, for the PSRs candidate class (Fig. \ref{fig:gradcam_stage1}), the model shifts its attention to the high-energy regime. The most relevant features are the detection significance in high-energy bands (\texttt{Sqrt\_TS\_Band\_6} and \texttt{Band\_7}). This confirms that the model utilizes the characteristic spectral hardness and stable GeV emission of PSRs as the primary discriminant, effectively ignoring the low-energy bands where AGNs dominate.
\end{itemize}

\subsection{Stage 2: YPs vs MSPs}
The second stage, which separates YPs from MSPs, presents a more subtle spectral challenge. The Grad-CAM results (Figure \ref{fig:gradcam_stage1}) highlight how the model resolves this:

\begin{itemize}
    \item YPs: The decision for YPs is overwhelmingly driven by the flux intensities in the highest energy bands (\texttt{Flux\_Band\_5} and \texttt{Flux\_Band\_6}, $3-30$\,GeV). This reflects the fact that young, energetic PSRs often exhibit spectral cutoffs at higher energies compared to the older, recycled MSP population, making them more prominent in the hard $\gamma$-ray band.
    
    \item MSPs: Interestingly, the MSPs classification relies heavily on uncertainty parameters (\texttt{Unc\_Flux\_Band}) across intermediate bands. This suggests that the model has learned to identify MSPs not just by their spectral shape, but by their characteristic faintness.
\end{itemize}

\section{Discussion}
\label{sec:discussion}

The classification of unassociated \textit{Fermi}-LAT sources has historically been challenged by an inherent reliance on spatial priors and the black-box nature of traditional machine learning algorithms. The results obtained by our hierarchical \textit{TabularResCNN} framework offer several critical advancements in both methodological robustness and astrophysical interpretation.

A primary achievement of this work is the successful decoupling of intrinsic emission physics from line-of-sight spatial biases. Previous studies \citep[e.g.,][]{2024MNRAS.527.1794Z} achieved high classification accuracies by heavily weighting spatial coordinates, effectively training models to recognize the Galactic plane rather than the intrinsic nature of the sources. By utilizing spectral shape, variability, and detection significance as sequential inputs, our 1D-CNN demonstrates that the intrinsic multi-band topology is sufficient for high-confidence classification. Consequently, our model is highly sensitive to the fainter, unassociated sources that comprise the bulk of the unknown 4FGL-DR4 population. This is reflected in the identification of 202 high-confidence pulsar candidates. By substituting the $\approx 0.5$ threshold with observationally motivated cuts (see Section \ref{sec:methods_thresholds}), we distilled this pure sample from a baseline of 1049 marginal detections, ensuring a highly robust addition to the current \textit{Fermi}-LAT pulsar census. Futhermore, on the training level we use weight loss function strategy instead of synthetic oversampling (eg. SMOTE) preventing the covariate shift that typically generate biases \citep{Malyshev_2025}.

The application of Grad-CAM reveals that the \textit{TabularResCNN} has autonomously learned fundamental high-energy astrophysics. For Stage 1 (AGNs vs. PSRs), the model's reliance on low-energy flux ($<300$\,MeV) and variability indices to identify AGNs. Conversely, the model identifies PSRs by focusing on the $>1$\,GeV bands, effectively tracing the characteristic exponential cutoff of pulsar magnetospheric emission. Interestingly, in Stage 2 (YPs vs. MSPs), the model leverages uncertainty parameters (\texttt{Unc\_Flux\_Band}) as key discriminants for MSPs. Rather than a statistical artifact, this represents a learned proxy for the characteristic lower signal-to-noise ratios of the MSPs population compared to the energetic YPs. This demonstrates that neural networks, when provided with a sensitivity mask (our Dual-Channel approach see section \ref{sec:features_engineering}), can extract physical meaning from observational limitations.

The spatial and dynamic properties of the newly classified candidates strongly validate the model's predictive power. Even without spatial features in the training phase, the strict confinement of the recovered YPs candidates to the Galactic plane confirms their identity as young objects born from recent core-collapse supernovae in the thin disk. In contrast, the broader vertical distribution of the MSPs candidates aligns with the expected kinematic diffusion of older systems (see Figure \ref{fig:latitude_dist}).

Other important point is the flux distribution (Figure \ref{fig:flux_distribtion}), our high-confidence sample of predicted PSRs exhibits relatively high gamma-ray fluxes, comparable to those of known 3PC PSRs. Because these candidates do not suffer from low photon statistics, their lack of detected \textit{Fermi}-LAT pulsations cannot be easily attributed to observational limitations. Rather, it indicates that our model is preferentially selecting a distinct underlying population of PSRs with intrinsic properties that hinder pulsation detection. This dichotomy is further evidenced by the $P-\dot{P}$ diagram (Figure \ref{fig:ppdot}), which clearly shows our candidates clustering in a separate parameter space from the standard \textit{Fermi}-LAT pulsar population.

While spatial coincidence with known ATNF PSRs provides strong physical validation, the high density of known PSRs along the Galactic plane ($|b| < 5^\circ$) introduces a non-negligible probability of chance associations. Consequently, a spatial cross-match in this crowded region yields inconclusive statistical significance when evaluated on a purely positional basis. Nevertheless, independent of whether all crossmatched candidates are true associations, it is evident from Figure \ref{fig:ppdot} that the YP and MSP populations remain distinctly isolated, accurately tracing their respective regimes in the $P-\dot{P}$ plane. Furthermore, as observed, the pulsar candidates generally exhibit a lower spin-down luminosity ($\dot{E}$) compared to the detected 3PC \textit{Fermi}-LAT sample. This likely explains why they have not been detected by the \textit{Fermi} satelite. To rigorously evaluate the predictive power of our model against this chance-coincidence background, we analyzed the relative enrichment of our candidate sample. We compared the fraction of successful ATNF cross-matches within our pulsar candidate list against the baseline cross-match rate of the general unassociated source population. Our analysis reveals a factor of $\sim3$ increase in the cross-match percentage when conditioning on our ML-classified pulsar candidates. This significant threefold enrichment demonstrates that the \textit{TabularResCNN} is not merely reproducing the spatial distribution of the Galactic disk, but is actively and successfully isolating the intrinsic spectral and temporal signatures of PSRs. The full statistical validation of which is detailed in Appendix \ref{app:statistical_validation}.

To further validate the robustness and generalization capabilities of our classification framework, we conducted a series of stringent tests addressing minority classes and spatial biases. First, we evaluated the model on unmodeled minority populations completely excluded from training. Theoretically, sources such as Supernova Remnants (SNRs) and gamma-ray binaries could be systematically confused with PSRs due to a curved spectra and a lack of significant time variability. However, empirical results demonstrate the high robustness of the learned feature space: the architecture resists this theoretical confusion by successfully isolating the unique spectral signature of genuine PSRs. Instead of being misclassified as PSRs, these minority classes are heavily driven away from the PSRs decision boundary, with the majority of SNRs and Pulsar Wind Nebulae (PWNe) being assigned to the AGNs class with very low PSR probabilities. This allows them to be naturally filtered out via high-confidence selection cuts. Second, rigorous spatial cross-validation tests across different Galactic latitudes ($|b| \le 15^\circ$ vs. $|b| > 15^\circ$) and longitudes ($l < 0^\circ$ vs. $l > 0^\circ$) demonstrated remarkable spatial invariance. This confirms that the architecture relies on intrinsic physical features rather than exploiting instrumental background shortcuts or flux uncertainty proxies, maintaining stable performance across the sky.

Building upon this demonstrated structural reliability, it is important to emphasize that the ultimate value of any predictive catalog lies in its utility for guiding future observations. The 100\% agreement between our high-confidence PSRs predictions and the recent independent blind-search discoveries by the FAST telescope (Table \ref{tab:fast_candidates}) validates the exceptional reliability of our pipeline. While Stage 2 exhibits a slightly lower accuracy (81.4\%) due to the intrinsic spectral overlap between some YPs and MSPs, the primary AGNs vs PSRs separation is highly robust. Most importantly, the 202 high-confidence pulsar candidates identified in this study provide a highly refined, bias-minimized target list for deep radio timing surveys, by optimizing our decision boundaries to maximize observational efficiency (Section \ref{sec:methods_thresholds}). Consequently, upcoming and current facilities, such as FAST, MeerKAT, and eventually the Square Kilometre Array Observatory (SKAO), can allocate time to this curated catalog with maximized expected discovery yields.

\section{Conclusions}
\label{sec:conclusions}

In this work, we develop a hierarchical deep learning framework, based on a custom 1D-CNN architecture (\textit{TabularResCNN}), to resolve the nature of unassociated $\gamma$-ray sources in the \textit{Fermi}-LAT catalog. By exploiting the intrinsic spectral and temporal topology of the data, our model transcends the limitations of traditional spatial classifiers, offering a robust probabilistic classification for 2563 previously unidentified sources. While our baseline model demonstrates exceptional separability (achieving a symmetric accuracy of 97.9\% in the primary AGNs vs PSRs discrimination), relying on generic mathematical metrics fails to account for the asymmetric costs of misclassification in observational astrophysics. By implementing a cost-sensitive threshold optimization, we explicitly aligned our decision boundaries with the operational realities of radio follow-up campaigns.

Prioritizing follow-up efficiency over absolute completeness, we systematically filtered marginal detections (Section \ref{sec:methods_thresholds}) to isolate a high-confidence catalog of 202 new pulsar candidates (comprising 166 YPs and 36 MSPs). This curated list increases the size of the 3PC catalog by more than 60\%, providing an optimized target list for upcoming surveys with radio facilities such as FAST, MeerKAT, and the SKAO. Applying the same technique to AGNs, we obtained a robust sample of 1136 new high-confidence AGNs candidates.

Crucially, they exhibit strong consistency with astrophysical priors. The spatial distribution of predicted YPs shows a strict confinement to the Galactic Plane ($|b| \approx 0^\circ$). Conversely, the predicted MSP population displays a broader vertical scale height, consistent with the kinematic heating expected from older PSR populations. Furthermore, the cross-matching of candidates with the ATNF database verifies that our model correctly maps the distinct regions of the $P-\dot{P}$ diagram, separating canonical PSRs from the millisecond regime without explicit kinematic inputs. The predictive power of this framework is empirically validated by its ability to correctly classify 5 out of 5 sources recently confirmed as PSRs by the FAST telescope (Table \ref{tab:fast_candidates}). This perfect agreement with independent radio discoveries underscores the utility of our catalog as a prioritized roadmap for future multi-wavelength campaigns. As next-generation facilities like the Square Kilometre Array Observatory (SKAO) and the Cherenkov Telescope Array Observatory (CTAO) come online.

\begin{acknowledgements}
C. Pozo González is hired under the Generation D initiative, promoted by Red.es, an organisation attached to the Ministry for Digital Transformation and the Civil Service, for the attraction and retention of talent through grants and training contracts, financed by the Recovery, Transformation and Resilience Plan through the European Union's Next Generation funds.

C.P.G., R.L.-C., J.M.-G. acknowledge financial support from the Spanish "Ministerio de Ciencia e Innovaci\'on" through grant PID2022-139117NB-C44 and grant CNS2023-144504. Authors also acknowledge financial support from the Severo Ochoa grant CEX2021-001131-S funded by MCIN/AEI/10.13039/501100011033. 
This research work was funded by the European Commission – NextGenerationEU, through Momentum CSIC Programme: Develop Your Digital Talent. 
J.M.-G. acknowledges financial support from the FPI-Severo Ochoa grant CEX2021-001131-S-20-6, PRE2022-103386 funded by MICIU/AEI/ 10.13039/501100011033 and ESF+.
EdOW acknowledges the support of DESY (Zeuthen), a member of the Helmholtz Association HGF.
We are grateful to Pablo Saz Parkinson for valuable discussions and comments during the paper drafting.
We gratefully acknowledge the anonymous referee for their valuable comments and suggestions, which significantly improved the quality of this manuscript.
\end{acknowledgements}

\bibliographystyle{bibtex/aa} 
\bibliography{bibtex/Bibliography} 

\begin{appendix}
\label{sec:implementation}
\section{Implementation}
The proposed framework was implemented using \texttt{PyTorch 2.x}, leveraging the \texttt{PyTorch Lightning} wrapper to standardize training loops and ensure checkpointing efficiency. Data ingestion and manipulation utilized the \texttt{Pandas} library, with the \textit{Fermi}-LAT 4FGL-DR4 catalog stored in the Apache Parquet columnar format to optimize I/O throughput during high-frequency data loading.

\subsection{Data Preprocessing and Input Encoding}
Prior to ingestion, numerical features were standardized using a \texttt{StandardScaler} ($z = (x - \mu) / \sigma$), with statistics ($\mu, \sigma$) computed strictly on the training partition to prevent look-ahead bias. Given the sparse nature of high-energy astronomical catalogs, we employed a Dual-Channel Input Strategy to handle missing values and upper limits without information loss:
\begin{itemize}
    \item Channel 1 (Features): Standardized numerical values, where missing entries are zero-filled.
    \item Channel 2 (Sensitivity Mask): A binary mask concatenated channel-wise, where $1$ denotes a valid measurement and $0$ indicates a missing value or upper limit. This explicitly allows the convolutional kernels to weigh feature significance based on instrumental sensitivity.
\end{itemize}

\subsection{Training Protocol and Optimization}
Optimization was performed using the Adam algorithm with an initial learning rate of $\eta = 10^{-3}$. We employed a \texttt{ReduceLROnPlateau} scheduler, decaying the learning rate by a factor of $\gamma = 0.5$ following 5 epochs of validation loss stagnation. To further improve generalization and calibration, we applied Label Smoothing ($\epsilon = 0.1$) to the Cross-Entropy loss function. 
Training was conducted with a batch size of 100 for a maximum of 50 epochs. An Early Stopping mechanism was configured to terminate training if validation loss failed to improve for 10 consecutive epochs, restoring the model weights associated with the highest validation accuracy.

\section{\textit{TabularResCNN} Architecture}
\label{app:architecture}

The \textit{TabularResCNN} is a 1-Dimensional Residual Convolutional Neural Network specifically designed to process tabular data with missing values. Its architecture can be conceptually divided into three stages: Input Formatting, Residual Feature Extraction, and Classification.

\subsection{Input Formatting}
The model takes two separate input tensors of size \texttt{(Batch\_Size, Num\_Features)}:
\begin{itemize}
    \item Values (\texttt{x}): The continuous or categorical representations of the features.
    \item Mask (\texttt{mask}): A binary placeholder mask indicating whether a feature was originally missing.
\end{itemize}

These are stacked along the channel dimension to form a 3D tensor of shape \texttt{(Batch\_Size, 2, Num\_Features)}, where the channels explicitly map the feature values and their presence.

\subsection{Residual Blocks (1D)}
The core feature extractor uses standard 1D Residual Blocks. Each block ensures gradient stability through a shortcut connection (skip connection). A standard block consists of:
\begin{enumerate}
    \item Conv1D (Kernel=3, Padding=1) $\rightarrow$ BatchNorm1D $\rightarrow$ ReLU
    \item Conv1D (Kernel=3, Padding=1) $\rightarrow$ BatchNorm1D
    \item Shortcut Addition $\rightarrow$ ReLU
\end{enumerate}

If spatial downsampling is required (stride=2) or the number of channels changes, the shortcut connection uses a $1\times1$ Conv1D followed by BatchNorm1D to match the dimensions.

\subsection{Global Average Pooling and Classification Head}
Following the residual stages, an Adaptive Global Average Pooling (\texttt{AdaptiveAvgPool1d}) collapses the spatial/feature dimension to 1. This removes strict sensitivity to the exact column ordering. The flattened tensor is then passed to a Multi-Layer Perceptron (MLP) for final classification.

\subsection{Architectural Summary Tables}

Table \ref{tab:macro_architecture} provides a layer-by-layer overview of the feature extraction backbone, while Table \ref{tab:classification_head} details the final fully connected classification head.

\begin{table*}
    \centering
    \caption{Macro Architecture Overview. Let $N$ be the number of initial tabular features, and $C$ the number of target classes.}
    \label{tab:macro_architecture}
    \small
    \begin{tabular}{llccl}
        \toprule
        \textbf{Stage} & \textbf{Layer / Module} & \textbf{Out. Channels} & \textbf{Out. Length} & \textbf{Stride / Details} \\
        \midrule
        \textbf{Input} & Tensor Stacking & 2 & $N$ & Concats value and mask mappings \\
        \textbf{Res 1} & Residual Block 1 & 16 & $N$ & Kernel=3, Stride=1, Padding=1 \\
        \textbf{Res 2} & Residual Block 2 & 32 & $\approx N/2$ & Kernel=3, Stride=2, Downsample=True \\
        \textbf{Res 3} & Residual Block 3 & 64 & $\approx N/4$ & Kernel=3, Stride=2, Downsample=True \\
        \textbf{Pool} & Global Average Pool & 64 & 1 & Reduces length dimension to 1 \\
        \textbf{Flatten}& Reshape / Flatten & - & - & Shape becomes \texttt{(Batch, 64)} \\
        \textbf{FC Head}& Linear Block & $C$ & - & See Table \ref{tab:classification_head} for details \\
        \bottomrule
    \end{tabular}
\end{table*}

\begin{table*}
    \centering
    \caption{Classification Head Details}
    \label{tab:classification_head}
    \small
    \begin{tabular}{llccl}
        \toprule
        \textbf{Layer} & \textbf{Type} & \textbf{Input Size} & \textbf{Output Size} & \textbf{Parameters / Notes} \\
        \midrule
        \texttt{fc.0} & Linear & 64 & 128 & Fully connected layer \\
        \texttt{fc.1} & ReLU & 128 & 128 & In-place activation \\
        \texttt{fc.2} & Dropout & 128 & 128 & Probability $p = 0.5$ \\
        \texttt{fc.3} & Linear & 128 & $C$ & Final logits projection \\
        \bottomrule
    \end{tabular}
\end{table*}

\section{Statistical Validation Methodologies}
\label{app:statistical_validation}

To robustly validate the physical association between our ML-classified candidates and known ATNF PSRs, we must account for the extreme surface density of sources along the Galactic plane. Both methods described below utilize a rigorous spatial cross-match algorithm that accounts for the positional uncertainty of Fermi sources. Rather than using simple radial distances, the Mahalanobis distance is computed by rotating the coordinates along the positional angle and scaling by the semi-major and semi-minor axes of the 68\% containment error ellipses ($\sigma_{a,b} \approx \mathrm{Conf}_{68} / 1.517$). The association probability is then exactly solved assuming a 2D Gaussian profile ($P = e^{-\Delta^2/2}$).

\subsection{Absolute Spatial Coincidence (Monte Carlo Analysis)}
\label{app:chance_coincidence}

Initially, we sought to demonstrate that the absolute number of associations between our ML candidates and ATNF PSRs significantly exceeded random spatial alignments. We employed a Monte Carlo approach, creating $N \geq 10,000$ mock implementations of the target population by randomizing their Galactic longitude ($\ell$) while keeping their Galactic latitude ($b$) intact, drawing from a Kernel Density Estimation (KDE) to preserve large-scale structures.

However, the results of this Monte Carlo analysis proved inconclusive. The underlying surface density of the ATNF catalog within the Galactic disk ($|b| < 5^\circ$) is so high that any randomized catalog confined to this region yields a massive number of chance coincidences. Consequently, evaluating the classification framework based purely on the absolute number of spatial cross-matches lacks discriminative statistical power, as the background noise of chance alignments overwhelms the signal.

\subsection{Pulsar Enrichment Analysis}
\label{app:enrichment_analysis}

Given the limitations of absolute spatial matching in the dense Galactic plane, we shifted our statistical approach to evaluate the relative predictive power of the model. We quantified the model's effectiveness by comparing the ATNF catalog association rate of our highest-confidence ML pulsar candidates against the baseline chance-association rate of the remaining unassociated sources.

To ensure a fair spatial comparison, we restricted our analysis to a bounding box containing the central 80\% of the ATNF pulsar population ($31.2^\circ \leq \ell \leq 310^\circ$ and $-3.9^\circ \leq b \leq 4.6^\circ$). This spatial filter isolated 347 unassociated 4FGL-DR4 sources located in the most densely populated region of the Galactic disk. 

We compared the fraction of successful ATNF cross-matches within our high-confidence ML candidates ($P_{ML} > 0.95$, $N=114$) against the baseline group of remaining unassociated sources ($N=233$). The baseline group exhibited a chance association rate of 7.73\% (18 matches). In stark contrast, the ML-selected candidate group achieved an association rate of 26.32\% (30 matches). 

This yields an enrichment factor of $3.4$, indicating that the model concentrates true PSRs more than three times better than a random spatial sampling of the exact same region. A Fisher's Exact Test confirms that this enrichment is highly statistically significant, returning a p-value of $p = 5.2 \times 10^{-6}$, which is equivalent to a $4.4\sigma$ confidence level. This strong evidence ($>3\sigma$) conclusively proves that, despite the overwhelming background density of the Galactic plane, the \textit{TabularResCNN} actively and successfully isolates the intrinsic spectral and temporal signatures of PSRs.

\section{Observational efficiency and the random selection baseline}
\label{app:efficiency}

To evaluate the operational impact of our CNN classification model on observational follow-up campaigns, we employ observational efficiency curves (analogous to lift or gain curves in the machine learning literature). In these representations, we benchmark the prioritization performance of our neural network against a control scenario denoted as random selection.

\subsection{Mathematical definition of random selection}

The random selection curve represents the expected discovery yield if telescope time were allocated to sources from the \textit{Fermi}-LAT unassociated catalog entirely at random, essentially simulating a blind survey with no prior spectral or morphological knowledge.

Mathematically, let $N$ be the total number of candidate sources in the catalog and $N_{\mathrm{target}}$ be the true number of sources belonging to the class of interest (e.g., PSRs and MSPs). The prior probability (or prevalence), $p$, of identifying the target class by pure chance is given by
\begin{equation}
    p = \frac{N_{\mathrm{target}}}{N}.
\end{equation}

Given an observational budget that permits targeting a subset of $K$ candidates, the expected number of successful discoveries under random selection, $\mathbb{E}[D_{\mathrm{rand}}]$, follows a hypergeometric distribution whose expected value is well approximated by a linear function:
\begin{equation}
    \mathbb{E}[D_{\mathrm{rand}}](K) = K \cdot p.
\end{equation}
This equation describes the linear baseline (labeled as random selection in our figures) whose slope is exactly the prevalence of the target class within the dataset.

\subsection{Quantifying observational gain}

The primary rationale for including this baseline is to strictly quantify the operational advantage provided by the deep learning model over a naive approach. We define the observational gain (or \textit{lift}) metric at a specific follow-up cutoff $K$ as the ratio between the true discoveries achieved by the CNN (when sorting sources by their predicted probability in descending order), $\mathbb{D}_{\mathrm{CNN}}(K)$, and the discoveries expected by chance:
\begin{equation}
    \mathrm{Lift}(K) = \frac{\mathbb{D}_{\mathrm{CNN}}(K)}{\mathbb{E}[D_{\mathrm{rand}}](K)}.
\end{equation}

\subsection{Physical and operational implications}

As demonstrated in our results, for restrictive observational thresholds (e.g., $K = 50$ targeted pointings), the CNN model yields a lift nearly an order of magnitude higher than random selection for the PSRs class. 

Physically, this implies that prioritizing targets based on the $P_{\mathrm{PSRs}}$ probability, extracted from underlying gamma-ray spatial and spectral signatures, rather than conducting unbiased catalog samplings, reduces the radio telescope integration time wasted on false positives (such as galactic background fluctuations or unresolved AGNs) by a factor equivalent to the lift. Consequently, the regime where the deviation between the model's yield curve and the random selection baseline is maximized defines the most cost-effective strategy for allocating resources in future follow-up campaigns.

\begin{figure}
    \centering
    \includegraphics[width=0.49\textwidth]{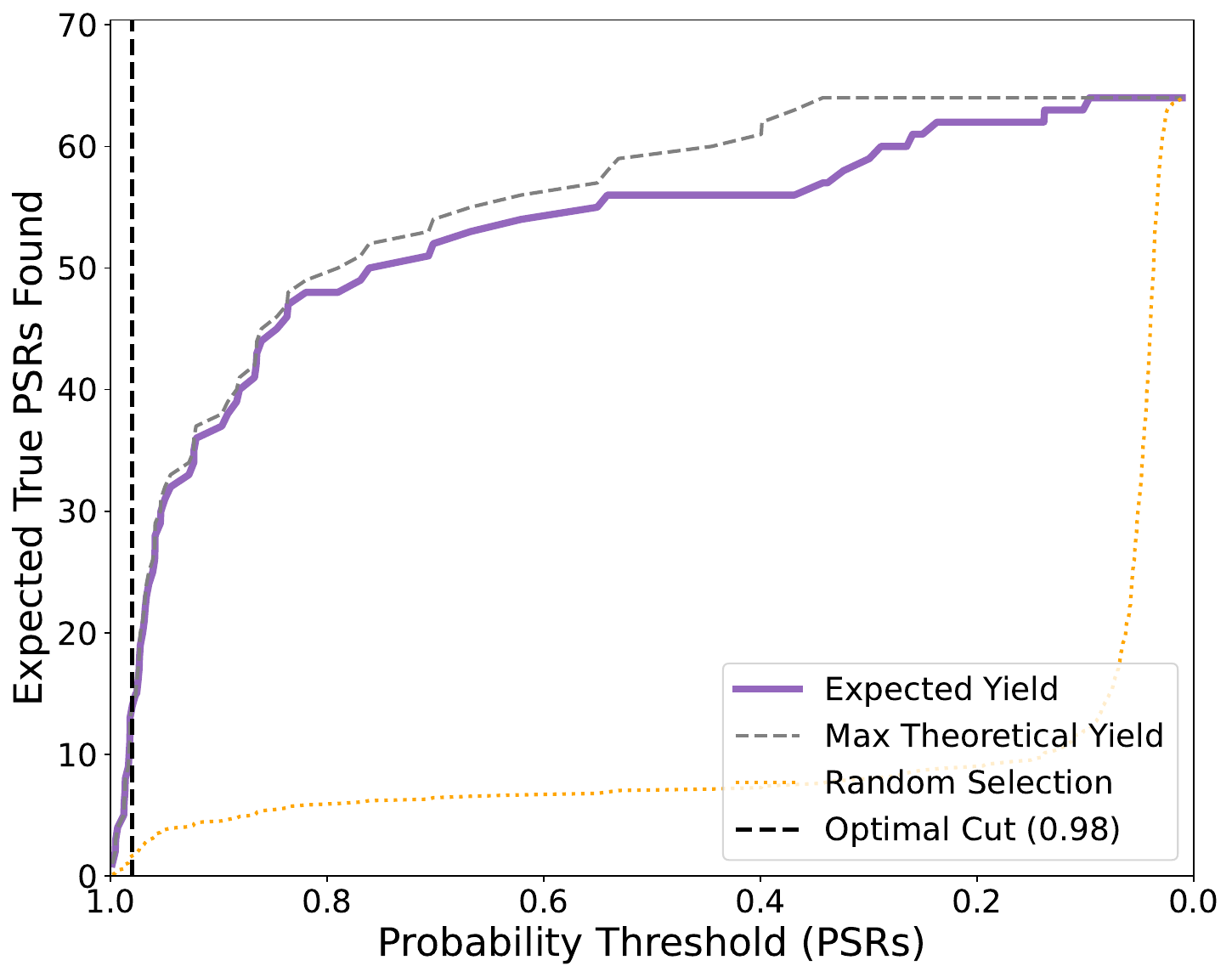}
    \label{fig:lift_psr}
    \includegraphics[width=0.49\textwidth]{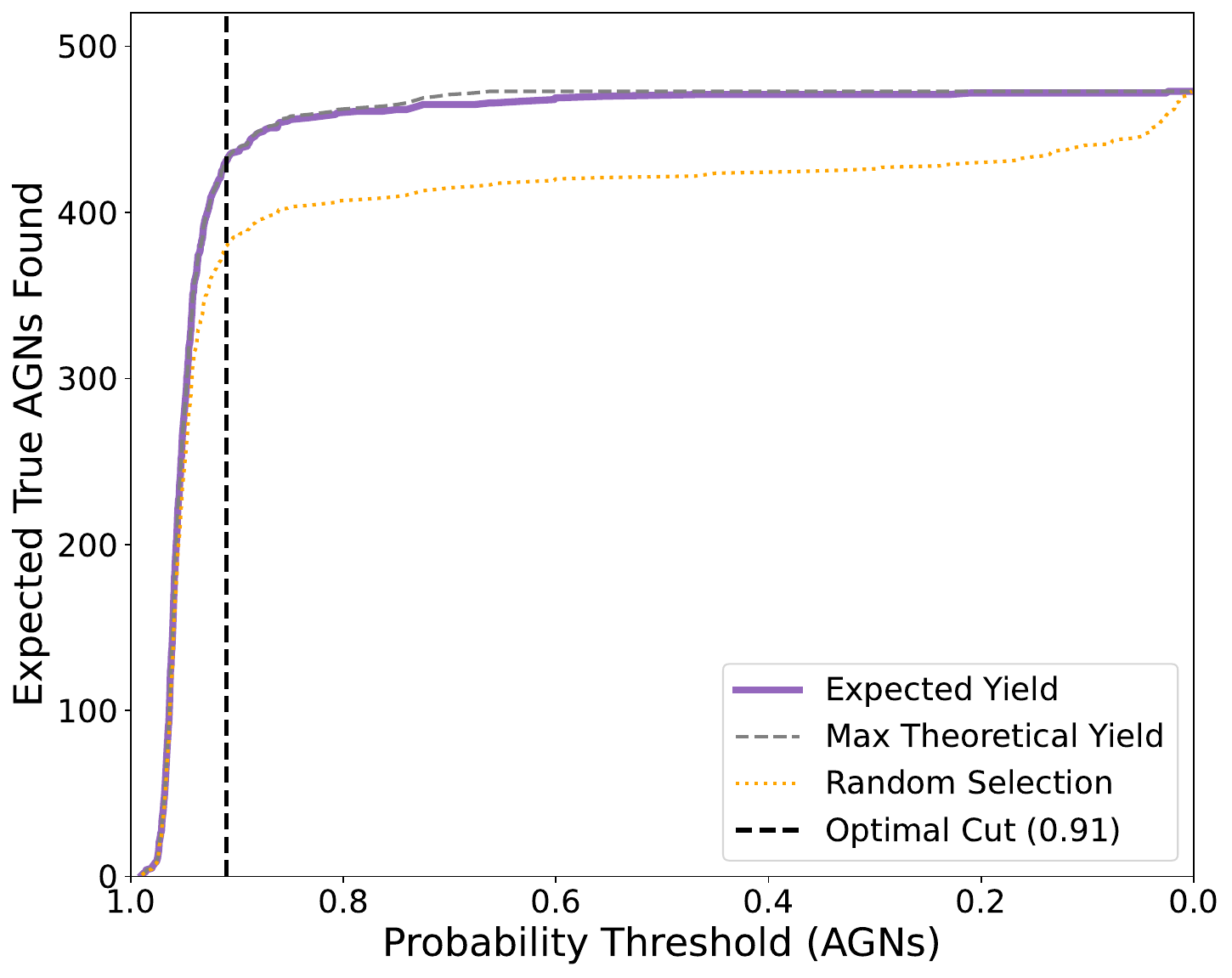}
    \label{fig:lift_agn}
    \caption{Observational efficiency (expected yield) as a function of the sweeping probability threshold for the unassociated \textit{Fermi}-LAT dataset. The solid purple line represents the model's expected discovery yield, the grey dashed line illustrates the theoretical maximum yield of the dataset, and the dotted orange line represents the baseline random selection. The vertical dashed black lines mark our physically motivated operational cuts: $P_{PSRs} = 0.96$ and $P_{AGNs} = 0.86$.}
    \label{fig:lift_curves}
\end{figure}

To visualize this operational advantage directly onto the decision boundaries applied in Section \ref{sec:results}, Figure \ref{fig:lift_curves} presents the expected discovery yield as a continuous function of the probability threshold for both the PSRs and AGNs classes.

Reading the x-axis from right to left (from $P=0.0$ to strict confidence at $P=1.0$), the curves illustrate the accumulation of true candidates. For the AGNs class (Figure \ref{fig:lift_curves} Botton panel), the expected yield closely traces the maximum theoretical yield even at exceptionally high thresholds. This confirms the high separability of extragalactic sources, justifying our strict operational cut at $P_{AGNs} = 0.86$ to guarantee maximum purity without sacrificing completeness.

Conversely, the PSRs curve (Figure \ref{fig:lift_curves} Upper panel) demonstrates the inherent difficulty of isolating faint PSRs from the Galactic background. While the model achieves a massive lift over random selection across all thresholds, enforcing a standard $0.5$ threshold would capture a high volume of candidates at the cost of a severely diluted observational efficiency (as the curve flattens significantly past $0.8$). The selected optimal cut of $P_{PSRs} = 0.96$ (vertical dashed line) is strategically positioned near the knee of the efficiency curve. This ensures that the generated catalog captures the most confident pulsar candidates before the expected yield begins to plateau, thereby maximizing the return on investment for subsequent radio telescope time.

\section{PSRs flux distribution}

\begin{figure}
    \centering
    \includegraphics[width=0.49\textwidth]{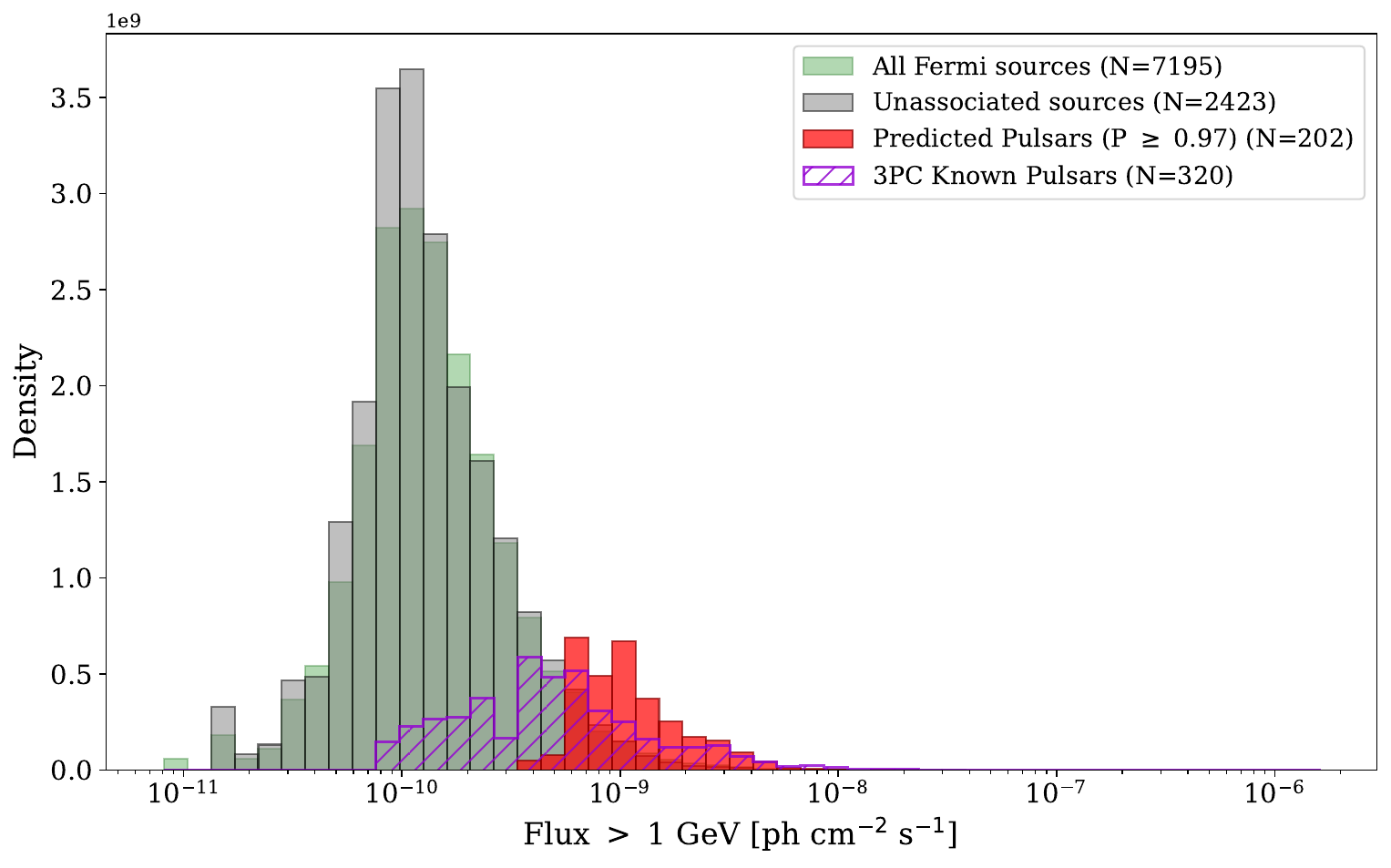}
    \label{fig:flux_distribtion}
    \caption{Density distribution of the gamma-ray flux ($> 1 \text{ GeV}$) for different source populations within the \textit{Fermi}-LAT catalog. The histogram compares all Fermi sources (green, $N=7195$), unassociated sources (grey, $N=2423$), our high-confidence predicted pulsar sample ($P \ge 0.97$, red, $N=202$), and known PSRs from the 3PC catalog (hatched purple, $N=320$). The predicted candidates generally exhibit higher fluxes than the bulk of the unassociated sources, closely matching the typical flux distribution of confirmed 3PC PSRs.}
\end{figure}

As shown in the flux distribution (Figure \ref{fig:flux_distribtion}), our high-confidence sample of predicted PSRs exhibits relatively high gamma-ray fluxes, comparable to those of known 3PC PSRs and significantly brighter than the bulk of unassociated Fermi sources. Given that these candidates do not suffer from low photon statistics, the lack of detected pulsations by \textit{Fermi}-LAT is unlikely to be an observational limitation due to low flux. Instead, this suggests that the absence of observed pulsed emission is driven by intrinsic properties of the sources themselves.

\end{appendix}

\end{document}